\documentclass[aps,amsmath,amssymb,prb,twocolumn,showpacs]{revtex4}
\usepackage{bm}
\usepackage{epsfig}

\newcommand{\bk}{{\bf k}}
\newcommand{\bq}{{\bf q}}

\newcommand{\br}{{\bf r}}
\newcommand{\bR}{{\bf R}}
\newcommand{\btr}{\tilde{\bf r}}
\newcommand{\bx}{{\bf x}}
\newcommand{\bA}{{\bf A}}
\newcommand{\bL}{{\bf L}}
\newcommand{\bU}{{\bf U}}
\newcommand{\bX}{{\bf X}}

\newcommand{\ct}{{\tilde{c}}}

\newcommand{\ud}{\textrm{d}}

\renewcommand{\H}{\hat{\mathcal{H}}}
\renewcommand{\P}{\mathcal{P}}
\newcommand{\D}{\mathcal{D}}
\renewcommand{\S}{\mathcal{S}}
\newcommand{\Z}{\mathcal{Z}}

\begin{document}

\title{Fermionic quantum criticality and the fractal nodal surface}

\author{Frank Kr{\"u}ger}
\author{Jan Zaanen}

\affiliation{Instituut-Lorentz, Universiteit Leiden, P. O. Box 9506, 2300 RA Leiden, The Netherlands}

\begin{abstract}
The complete lack of theoretical understanding of the quantum critical states found in the heavy fermion metals 
and the normal states of the high-T$_c$ superconductors is routed in deep fundamental problem of 
condensed matter physics: the infamous minus signs associated with Fermi-Dirac statistics render  
the path integral non-probabilistic and do not allow to establish a connection with critical phenomena
in classical systems. Using Ceperley's constrained path-integral formalism 
we demonstrate that the workings of scale invariance and Fermi-Dirac statistics can be reconciled.  
The latter is self-consistently translated into a geometrical constraint structure. We prove that this 
"nodal hypersurface" encodes the scales of the Fermi liquid and turns fractal when the system becomes 
quantum critical. To illustrate this we calculate nodal surfaces and electron momentum distributions of 
Feynman backflow wave functions and indeed find that with increasing backflow strength the 
quasiparticle mass gradually increases, to diverge when the nodal structure becomes fractal. Such a
collapse of a Fermi liquid at a critical point has been observed in the heavy-fermion intermetallics in a
spectacular fashion.
\end{abstract}

\maketitle

\section{Introduction}
Not long ago, it was taken as physical law that macroscopic systems formed from fermions are
Fermi liquids, behaving in a scaling sense as the non-interacting Fermi gas characterized by the 
Fermi degeneracy scale. This has changed drastically in recent times by the discovery that in some 
metals the system of electrons behaves very differently. Clear cut examples are  the 'quantum critical' 
metallic states found at quantum phase transitions in heavy fermion compounds\cite{schofield,coleman,si,
stewart,custers,paschen}, while there are indications that the metallic states found in the cuprate high-T$_c$
superconductors are of a similar kind.\cite{coleman,takagi,sachdev,marel,orenstein,zhu,zaanen,zaanen2,
varma97} 
By tuning a zero temperature control parameter (like pressure, magnetic field, density) one 
encounters Fermi liquids with different Fermi surfaces, and the quantum critical regime is found at the zero 
temperature phase transition where one metal turns into the other\cite{custers,paschen}. 
Physical properties in this quantum critical regime are controlled by powerlaws, indicating that the system 
has become scale invariant, in analogy with both thermal phase transitions and the quantum phase transitions 
in bosonic systems. Albeit rooted in quantum statistics, the Fermi energy is a scale and has therefore to vanish 
in the quantum critical regime. This is confirmed in a spectacular fashion in the heavy-fermion systems: 
the mass of the Landau quasiparticles in the Fermi liquids on both sides of the transition should 
be inversely proportional to the Fermi energy, and this mass is found to diverge to infinity in the 
quantum critical regime.\cite{custers}

How to think about a fermion liquid without Fermi energy? The complete lack of success in
understanding the above phenomena is caused by a deep and general methodological problem
in many-particle quantum physics.\cite{zaanen08} For bosonic problems one can employ the powerful path 
integral methods of quantum field theory, directly relating e.g. the quantum critical state to the well understood
statistical physics of classical phase transitions. For fermionic systems this alley is blocked by the 
infamous minus-sign problem rendering the path integral non-probabilistic. In fact the mathematics is
as bad as it can be: Troyer and Wiese\cite{Troyer+05} showed recently that the sign problem falls in the
mathematical complexity class "NP hard", and the Clay Mathematics Institute has put one of its 7 one 
million dollar prizes on the proof that such problems cannot be solved in polynomial time.

Some time ago, Ceperley\cite{ceperley} discovered an alternative representation for the fermionic path 
integral that does not solve the minus-sign problem in a mathematical sense but has as virtue that the fermionic 
statistics is coded in a more manageable way: the 'constrained' fermionic path integral. In this framework, the 
minus signs associated with Fermi-Dirac statistics are self-consistently translated into a geometrical constraint 
structure (the nodal hypersurface) acting on a residual bosonic dynamics. Although this nodal surface which 
contains all the data associated with the differences between bosonic and fermionic matter is a priori not known
for an interacting fermion problem, in the scaling limit only its average and global properties should matter. 
Henceforth, it should be possible in principle to classify all forms of fermionic matter in a phenomenological way 
by classifying the average geometrical- and topological properties of the constraint structure to subsequently 
use this data as an input to solve the resulting bosonic path integral problem. This procedure is supposedly a 
unique extension of the Ginzburg-Landau-Wilson paradigm for bosonic matter to fermionic matter. Employing 
the Ceperley path integral, in this paper we deliver proof of principle that fermion statistics and emergent scale
invariance underlying the critical state can be reconciled.

This paper is organized as follows. In Section \ref{sec.constrained} we introduce the Ceperley path integral
and explain the notion of the "nodal hypersurface". As one anticipates, it has to be that the scales of the Fermi 
liquid are encoded in the nodal surface since the residual bosonic dynamics cannot possibly generate these scales 
by itself. In Section \ref{sec.geometry} we indeed establish a one-to-one correspondence
between the Fermi degeneracy scale and an average nodal pocket dimension and demonstrate that in order for
a fermionic system to become critical the nodal surface has to turn into a scale invariant fractal. This should be 
regarded as our most important result which is further illustrated in the remainder of this paper.
In Section \ref{sec.feynman} we introduce the concept of Feynman and Cohen\cite{feynman} of incorporating hydrodynamical 
backflow effects in a quantum mechanical wave function and show that fermionic backflow wave functions describe
many-particle states characterized by a hierarchy of increasing number of particle correlations. The nodal structures
of such wave functions are investigated in Section \ref{sec.backflow}.
We find that their nodal surfaces change drastically with increasing backflow strength and 
turn into a fractal when the backflow becomes hydrodynamical, involving a macroscopic number of particles. 
A detailed fractal analysis is provided in Section \ref{sec.fractal}. In Section \ref{sec.mass} we perform Monte-Carlo 
calculations of the momentum distribution to extract the quasiparticle effective mass as a function of backflow strength. 
We find that the effective mass diverges exactly at the point where the nodal surface turns into a fractal. Finally, 
in Section \ref{sec.disc} our results are summarized and discussed.

\section{Constrained path integral and nodal surface}
\label{sec.constrained}

The fermion sign problem becomes apparent when expressing the many body density 
matrix $\rho_F(\bR,\bR';\hbar\beta)$  with $\bR=(\br_1,\ldots,\br_N)$ the 
position in $dN$ dimensional configuration space ($d$ the spacial dimension and $N$ the number of 
particles),  and $\beta=1/(k_B T)$ the inverse temperature, as a path 
integral over worldlines $\{\bR_\tau\}$ in imaginary time $\tau$ ($0\leq \tau\leq\hbar\beta$), weighted by an action 
$\S[\bR_\tau]$, 

\begin{eqnarray}
\rho_F(\bR,\bR';\hbar\beta) & = &\frac{1}{N!}\sum_\P (-1)^p\int_{\bR\to\P\bR'}\D\bR_\tau e^{-\S[\bR_\tau]/\hbar}\nonumber\\
\S[\bR_\tau] & = & \int_0^{\hbar\beta}\ud\tau\left\{\frac m2 \dot{\bR}^2_\tau+V(\bR_\tau)\right\},
\label{PI_signful}
\end{eqnarray}
where the sum over all possible $N!$ particle permutations $\P$ accounts for the indistinguishability of the particles 
and the alternating sign imposes the Fermi-Dirac statistics. Here $p=\textrm{par}(\mathcal{P})$ denotes the parity 
of the permutation, even permutations enter with a positive, odd permutations with a negative sign. 
The term $V(\bR)$ is a short hand notation for both
external potentials and particle interactions. For simplicity, we have considered spinless fermions.  
The partition function is obtained as a trace over the diagonal 
elements of the density matrix, $\Z_N(\beta)=\int\ud\bR\rho_F(\bR,\bR;\hbar\beta)$ corresponding to worldlines 
returning to their starting place or a permutation of it. For bosons where the minus signs are absent the 
partition function can be viewed as a classical one describing an ensemble of interacting cross-linked 
ringpolymers. However, the fermionic minus signs make this probabilistic interpretation impossible.

Some time ago, Ceperley proved\cite{ceperley}
that the fermionic density matrix can be calculated as a path integral analogous to Eq. (\ref{PI_signful}) but summing 
only over worldlines that do not cross the nodes of the density matrix itself which define for each given initial point 
$\bR_0$ and inverse temperature $\beta$ a $(dN-1)$-dimensional hypersurface in $dN$-dimensional configuration space, 

\begin{equation}
\Omega_{\bR_0,\beta}:=\{\bR|\rho_F(\bR_0,\bR;\hbar\beta)=0\}.
\end{equation}
Those hypersufaces act as infinite potential barriers 
allowing only for node avoiding worldines $\bR_\tau$ with $\rho_F(\bR,\bR_\tau;\tau)\neq 0$ for $0\leq\tau\leq\hbar\beta$. 

To calculate the partition function we have to integrate over the diagonal density matrix $\rho_F(\bR,\bR;\hbar\beta)$ which
is obtained as a path integral over all worldline configuration $\bR \to{\cal P}\bR$ which do not cross the nodal surface on any 
time slice and belong to the reach

\begin{equation}
\Gamma_\beta(\bR) = \{ \gamma : \bR \rightarrow \bR' | \rho_F (\bR, \bR (\tau); \tau ) \neq 0 \}.
\label{reach}
\end{equation}
Because of the anti-symmetry of the fermionic density matrix under particle permutations ${\cal P}$, 

\begin{eqnarray}
\rho_F(\bR,{\cal P}\bR;\hbar\beta) & = & \rho_F({\cal P}\bR,\bR;\hbar\beta)\nonumber\\
 & = & (-1)^p\rho_F(\bR,\bR;\hbar\beta),
\end{eqnarray}
all worldline configurations corresponding to odd permutations have to cross a node an odd number of times and are therefore completely removed from the partition function. They are exactly cancelled out by all node crossing even permutations and 
we are left with an ensemble of all node-avoiding worldline configurations corresponding to even permutations,

\begin{equation}
\rho_F (\bR,\bR;\hbar\beta )=\frac{1}{N!}\sum_{{\cal P}, \textrm{even}}\int_{\gamma: \bR \to{\cal P}\bR}^{\gamma \in \Gamma_\beta(\bR)}
{\cal D}\bR_\tau e^{-\mathcal{S}[\bR]/\hbar}.  
\label{constpath} 
\end{equation}

Remarkably, this representation of an arbitrary fermion problem is not suffering from the 'negative probabilities' 
of the standard formulation. Surely, one cannot negotiate with the NP-hardness of the fermion problem and Ceperley's 
path integral is not solving this problem in a mathematical sense. However, the negative signs are transformed away 
at the expense of a structure of constraints limiting the Boltzmannian sum over worldline configurations. These 
constraints in turn can be related to a geometrical manifold embedded in configuration space: the 'reach', which is 
determined by the nodal hypersurfaces of the fermion density matrix. This reach should be computed self-consistently: 
it is governed by the constrained path integral that needs itself the reach to be computed. 

So far, this self consistent reformulation of the path integral has only been used to study Fermi liquids like
helium-3\cite{ceperley2} and many body hydrogen\cite{pierleoni,magro} quantitatively within path-integral 
Monte-Carlo simulations fixing the nodal constraints with a free-particle density matrix. At a first glance, this might
seem to be a very crude approximation, however, in a scaling sense the Fermi liquid can be viewed as a gas of weakly
interacting Landau quasiparticles. As we will see in the next section, imposing the nodal constraint structure of 
non-interacting fermions on the residual bosonic dynamics means nothing but taking Landau's Fermi-liquid paradigm 
for granted.

Let us further inspect the nodal hypersurface and the workings of the constrained path integral. Since the 
fermionic density matrix is odd under particle exchanges $\mathcal{P}_{ij}$,

\begin{equation}
\rho_F(\bR_0,\mathcal{P}_{ij}\bR;\hbar\beta)=-\rho_F(\bR_0,\bR;\hbar\beta),
\end{equation}
the density matrix is zero if two fermions are at the same position, $\br_i=\br_j$, irrespective of the temperature, 
the reference point $\bR_0$, and the interactions between the particles. Therefore, the Pauli surface

\begin{equation}
P  =  \bigcup_{i,j}^{i\neq j}  \{\bR|\br_i=\br_j\}
\end{equation}
is always a submanifold of the nodal surface $\Omega_{\bR_0,\beta}$, whereas the dimensionality of the Pauli surface
is $\textrm{dim}P=Nd-d$.

In terms of a complete set of fermionic eigenfunctions $\Psi_\alpha(\bR)$ with eigenvalues $E_\alpha$ the fermionic 
density matrix is given by
\begin{equation}
\rho_F (\bR_0,\bR;\hbar\beta )=\sum_\alpha e^{-\beta E_\alpha}\Psi^*_\alpha(\bR_0)\Psi_\alpha(\bR),
\end{equation}
which, in case of a non-degenerate ground state $\Psi_0(\bR)$, converges in the limit $T\to 0$ to 
$\rho_F({\bf R_0}, {\bf R}; \hbar\beta = \infty) = \Psi^* ({\bf R_0}) \Psi ({\bf R})$. Therefore, the nodal surface 
$\Omega_{\bR_0,\beta}$ of the finite temperature density matrix becomes independent of the reference point $\bR_0$ 
in the zero-temperature limit and converges to the nodal surface of the ground-state 
wave function, $\Omega:=\{\bR|\Psi_0(\bR) =0\}$.

\begin{figure}
\begin{centering}
\includegraphics[width=\linewidth]{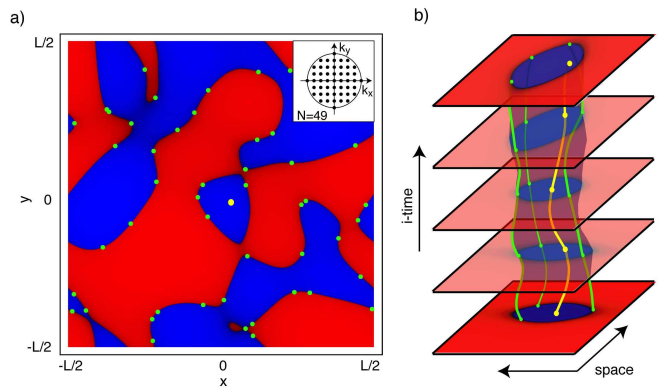}
\end{centering}
\caption{a) Cut through the nodal hypersurface of the ground-state wave function of $N=49$ free, spinless fermions in a two-dimensional box with periodic boundary conditions. The set of momentum states corresponding to a non-degenerate ground state is shown in the inset. The cut is obtained by fixing $N-1$ fermions at random positions (green dots) and moving the remaining particle (yellow dot) over the system. The nodal surface cut is given by the interface between red and blue regions corresponding to negative and positive values of the wave function, respectively, whereas absolute values are encoded in the color shading. The nodal lines  connect the $N-1$ fixed particles since the $(Nd-d)$-dimensional Pauli surface is a lower dimensional submanifold of the 
$(Nd-1)$-dimensional nodal hypersurface. b) Sketch of an allowed world-line configuration contributing to the Ceperley path integral.
On every time slice $\tau$ one particular particle (yellow dot) sees a nodal surface determined by the positions of the other particles 
on this time slice. Due to the meanderings of the worldlines of the other $N-1$ particles (green) the nodal surfaces form a 
'tent'  in space-time attached to the particle worldlines. This nodal 'tent' acts as a hard core boundary for the worldline of the remaining particle (yellow line).}
\label{freeandtent}
\end{figure}

In Fig. (\ref{freeandtent}a) a random cut through the nodal hypersurface of the ground-state wavefunction $\Psi(\br_1,
\ldots,\br_N) = \mathcal{N}\det\left(e^{i\bk_i\br_j}\right)_{i,j=1,\ldots,N}$ of $N=49$ spinless fermions in a two-dimensional
periodic box is shown. This particular particle number corresponds to a set of momenta $\bk_1\ldots,\bk_N$ on a grid 
$\Delta k=2\pi/L$ ($L$ the linear dimension of the box) forming a closed shell in momentum space and therefore   
to a non-degenerate ground state (see inset of Fig. (\ref{freeandtent}a)). The cut is obtained by fixing $N-1$ particles at random 
positions and tracking down the nodes of the wave function moving the remaining particle over the system. The algorithms used to find the nodes of free fermion and Feynman backflow wave functions studied later on is described in detail in appendix \ref{a.nodes}.
Since the Pauli surface
is a lower dimensional submanifold of the nodal surface the fixed $N-1$ particles are located on the nodal lines. We find that 
the nodes are very smooth forming pockets of the order of the average inter-particle spacing $r_s$. An investigation 
of the nodal structures of the finite-temperature density matrix\cite{ceperley} shows that this observation holds at any temperature
and that the nodal surfaces smoothly approach the ground-state nodes in the limit $T\to 0$.

The workings of the constrained path integral is sketched in Fig. (\ref{freeandtent}). On every time slice $\tau$ one particular 
particle $\br_1$ sees the nodal constraint structure $\Omega_{\bR_0,\beta}:=\{\bR|\rho(\bR_0,\bR;\tau)=0\}$ determined by the 
positions $\br_2(\tau),\ldots,\br_N(\tau)$ of the $N-1$ other particles on this time slice. Due to the meanderings of the worldlines 
of the $N-1$ particles the nodal surfaces form a 'tent' in space-time attached to the particle worldlines since the Pauli surface is always
a lower-dimensional nodal hypersurface. Since $\rho_F (\bR, \bR (\tau); \tau ) \neq 0$ for all $0\leq\tau\leq \hbar\beta$ this
nodal 'tent' acts as a hardcore boundary for the world line of the particle $\br_1$ and the particle is not allowed to penetrate or collide
with the 'tent'. A world line configuration as shown in Fig. (\ref{freeandtent}) does not violate the constraints and contributes to the Ceperley  path integral.

In a recent tutorial paper\cite{zaanen+08} the fermion sign problem has been studied within both the conventional signful
and the Ceperley path integral. It turns out that even for the free Fermi gas for which every student in physics knows the 
canonical solution, the constrained path integral turns into a highly nontrivial affair. Remarkably, in momentum space the 
constrained path integral directly leads to a one-to-one correspondence between the Fermi gas and a system of classical
atoms forming a Mott insulating state in the presence of a commensurate optical lattice of infinite strength, living in a harmonic
potential trap of finite strength.\cite{zaanen+08} This analogy is literal and the only oddity is that we are talking about an optical
lattice system in momentum space. We immediately rediscover our canonical picture of the Fermi gas, simply because the 
dynamics of the world lines becomes trivial due to the conservation of single-particle momentum. However, the workings of the
nodal constraints in the real-space formulation remain to a great extend puzzling.\cite{zaanen+08} 

Let us start with the case of free fermions in one space dimension, where the physics of quantum matter can be regarded as
completely understood.\cite{giamarchi} The deep reason is that quantum statistics has no physical meaning in 1+1D, and it is
always possible to find a representation where the sign structure drops out completely. It is instructive to find out how this is processed by the Ceperley path integral. The special status of the one-dimensional case becomes immediately clear since the dimensions
of the Pauli surface and the nodal hypersurface coincide and the two manifold become the same. In this situation it becomes
quite easy to read the reach. Start out with a reference point $\bR_0=(x_1,x_2,\ldots,x_N)$  ordering the particles for
instance like $x_1<x_2<\cdots<x_N$. 'Spread out' this configuration in terms of world lines meandering along the imaginary time direction and the Pauli-hypersurface reach tells that only configurations are allowed where these world lines never cross each
other at any imaginary time. Therefore, the ordering of the particles is preserved on all times and we only have to consider 
world-line configurations where every particle returns to its starting position. The particles become effectively distinguishable.
A more abstract way of saying this is that the Pauli surface (=nodal surface) divides the $N$-dimensional configuration space 
into $N!$ disconnected nodal cells, each corresponding to a certain ordering of the particles. We have seen that the Ceperley
path integral for the one-dimensional Fermi gas is equivalent to the one for hard-core bosons in 1+1 dimensions or from a statistical physics point of view to the problem of an ensemble of polymers with only steric, hard-core interactions in 2 dimensions.
This 'Pokrovsky-Talapov' problem\cite{pokrovsky} is surely a very serious statistical physics problem since the constraints 
correspond with infinitely strong delta function interactions and accordingly everything is about entropic interactions and order-out-of-disorder physics. Remarkably, the one dimensional fermion story can be completely understood from this radically statistical
physics viewpoint\cite{zaanen00,mukhin+01} by using a 'self-consistent phonon' method discovered by Helfrich\cite{helfrich} to
deal with the entropic interactions associated with biological (extrinsic curvature) membranes. 

Obviously, in $d\ge 2$ where the Pauli surface is a lower dimensional submanifold of the nodal 
hypersurface the self-consistency problem inherent to the constrained path integral cannot be resolved like in the 
one-dimensional case. This is the reason why a general bosonization procedure is lacking in higher dimensions. 
Although the Ceperley path integral has a much richer structure in higher dimensions it is surely the case that the higher
dimensional Fermi liquids have to know about the 'entropic dynamics' characteristic for the (1+1)-dimensional case.
However, a generalization of the Helfrich construction to deal with the steric interactions with nodal 'tent' in a self-consistent
way is lacking. Moreover, in contrast to the one-dimensional case, in $d\ge 2$ world line configurations corresponding to even permutations are a-priori not ruled out by the constraint structure. Recently, Mitas demonstrated\cite{mitas06a,mitas06b} 
that in the case of free spinless fermions the nodal surface has a minimal tiling property: it divides the 
configuration space into two nodal cells only. Therefore, it is alway possible to find a continuous path $\bR\to\mathcal{P}_{\textrm{even}}\bR$ not encountering a node. In other words,
all even permutations are on the reach $\Gamma_\beta(\bR)$ and contribute to the constrained path integral.

Phenomenologically the Ceperley path integral can be viewed as a bosonic dynamics subject to a geometrical 
constraint structure and therefore as a statistical physics problem. However, it is a highly non trivial question how to 
reconstruct the free $d$-dimensional Fermi gas within this framework using the real space representation (\ref{constpath}) 
and only the one-dimensional case can be regarded as fully understood.\cite{zaanen00,mukhin+01} From the 
canonical picture we know that the Fermi gas is characterized by a sharp Fermi surface at zero temperature and that the thermodynamics at low temperatures is governed by particle-hole excitations in
the vicinity of the Fermi surface leading for instance to a linear specific heat $C(T)\sim k_BT/E_F$ for $T\ll E_F$, irrespective
of the spatial dimension. This is surely unconventional for a system of interacting bosons where the spatial dimension enters
the exponents of low-temperature expansions in a natural way. Recently, it was conjectured\cite{zaanen+08} that the
nodal surface constraints act in a highly nonlocal way leading to an effective reduction of the dimensionality in a way that the systems behave qualitatively like soft-core bosons in 1+1D! Such a 'holographic' principle would also explain 
why the nodal-surface constraints prevent the system from undergoing a Bose condensation at finite temperatures and 
might relate the emergence of a sharp Fermi surface with a condensation exactly at $T=0$.

\section{Geometrical view on Fermi-Dirac statistics and scale invariance}
\label{sec.geometry}

The constrained path integral  is a precise reformulation of the sign-full path-integral in terms 
of an effective bosonic dynamics subject to a geometrical constraint structure and therefore leads the way 
to a probabilistic, statistical physics interpretation of fermionic systems since the sign structure is absorbed in the 
nodal hypersurface. A fermionic state like the Fermi liquid is characterized by scales, the Fermi-energy 
$E_F$ and momentum $k_F$. These scales are alien to any bosonic system. It has to be that these 
scales are uniquely encoded in the nodal structure, since the residual bosonic system cannot possibly generate 
these scales by itself.    

In the previous section, we have seen that in dimensions $d\ge 2$ the Pauli surface is a lower dimensional sub-
manifold of the nodal hypersurface, irrespective of temperature or the form of interactions. For free fermions we 
find the nodal surface to smoothly connect the lower dimensional Pauli surface both at zero temperature
(see Fig. (\ref{freeandtent}a)) and in the finite temperature case.\cite{ceperley} Therefore, the corresponding nodal
structure is characterized by a scale, an average nodal spacing of the order of the average inter-particle spacing $r_s$. 
This is clearly seen in the two-dimensional nodal surface cut shown in Fig. (\ref{freeandtent}a): moving one particular 
particle over the two-dimensional box we find nodal lines smoothly connecting the $N-1$ other particles, forming pockets
with a linear dimension of the order of $r_s$.

The presence of this scale in the nodal surface can be also deduced by a different argument.\cite{ceperley} From the 
reduced one body density matrix which is simply the Fourier transform of the
single-particle momentum distribution, $n(\br)=\int_{\bk}e^{i\bk\br}n_{\bk}$, given in the limit $T\to 0$ by

\begin{equation}
n(\br)=\int\ud\bR\Psi^*(\br_1,\br_2,\ldots,\br_N)\Psi(\br_1+\br,\br_2,\ldots,\br_N),
\label{nr}
\end{equation}
one obtains an estimate for the nodal spacing. The zeros of $n(r)$ correspond to the average displacement of 
a particle to be found on a node. For the Fermi gas the momentum distribution in the limit $N\to\infty$ simply turns into a step function $n_{\bk}=\Theta(|\bk|-k_F)$ with $\Theta(x)=1$ for $x\le0$ and $\Theta(x)=1$ for $x>0$.  This leads to the Fourier transform 

\begin{equation}
n(\br)=c_d (k_F r)^{-d/2}J_{d/2}(k_F r),
\label{nrfree}
\end{equation}
where $c_d$ denotes a constant depending on the spatial dimension and $J_{d/2}$ a Bessel function of the first kind. Asymptotically,

\begin{equation}
n(\br)\sim(k_F r)^{-(d+1)/2}\cos\left(k_F r-\frac{d+1}{4}\pi\right),
\end{equation}
signaling a long-range periodicity in the nodal structure with an average nodal spacing $r_n\sim k_F^{-1}\sim r_s$. This
asymptotic behavior is also generic for the Fermi liquid characterized by a discontinuity in $n(k)$ at the Fermi wave-vector 
$k_F$.
 
From the constrained path integral a one-to-one correspondence between the existence of a scale in the nodal 
hypersurface and the Fermi energy $E_F$ can be established by a simple scaling argument. Let us first assume that the nodal hypersurface is characterized by an average nodal pocket dimension of the order of the inter-particle spacing $r_s$ 
characteristic for the Fermi liquid. An allowed world line configuration is sketched in Fig. (\ref{freeandtent}b) where we follow 
the time evolution of the nodal constraint structure seen by one particular particle. In the Ceperley path integral the world-line configurations $\bR_\tau$ are constrained by the reach determined by the condition that the world lines are not allowed to cross the 
nodal hypersurface of the density matrix itself at any time. From the perspective of one particular particle this means that
the particle has to stay in its nodal pocket at all times $0\le\tau\le\hbar\beta$. Due to the meanderings of the world lines of the
other particles the nodal structure seen by the particle fluctuates in time leading in a time continuum limit to the picture of a nodal
surface 'tent' which hangs in space-time and acts as a hard-core boundary for the particle. Since the Pauli surface is at all times a lower dimensional submanifold of the nodal hypersurface the world lines of the other particles act as twisted 'tent-sticks' on which the 'tent' is hanging. 

Every particle has to stay within the nodal 'tent' which is governed by the dynamics of the other particles. At a timescale $\tau_c$ 
when the average square displacement

\begin{equation} 
l^2(\tau) =  \langle[\br_i(\tau)-\br_i(0)]^2\rangle = 2d\frac{\hbar}{2m}\tau
\end{equation}
of the worldlines becomes of the order of the average nodal spacing, $l(\tau_c)= r_n$, the particles start to 
collide with the nodal tent. This leads to an average collision time $\tau_c= (2d)^{-1}(2m/\hbar)r_n^2$ corresponding to
an energy scale 

\begin{equation}
E_c=\frac{\hbar}{\tau_c}  = 2d\frac{\hbar^2}{2m}r_n^{-2}\simeq  \frac{2\pi d}{\Gamma^{2/d}(\frac d2 +1)}\frac{\hbar^2}{2m}n^{2/d},
\label{Ec}
\end{equation}
where in the last step we have used that the nodal spacing $r_n$ is of the order of the inter-particle spacing $r_s$, $r_n\simeq r_s$ and introduced the particle density $n=N/V= r_s^{-d}/K_d$ with $K_d=\pi^{d/2}/\Gamma(d/2+1)$ the volume of the $d$-dimensional unit sphere. We recognize immediately that Eq. (\ref{Ec}) is just the expression for the Fermi energy in $d$ dimensions, $E_F\simeq E_c$. In the case of free fermions we can even be more quantitative and estimate the average nodal pocket dimension $r_n$ 
as the first zero of the reduced one-body density matrix $n(r)$ (\ref{nrfree}) which in $d$ dimensions is given by $r_n=z_d/k_F$
with $z_d$ the first zero of the Bessel function $J_{d/2}(z)$. This yields $E_c=\alpha_d E_F$ with $\alpha_2\approx 0.27$,
$\alpha_3\approx 0.30$ in two and three dimensions, respectively.

From the above scaling argument we have learned that an average nodal pocket dimension is dynamically related to a typical 
timescale $\tau_c$ on which the particles feel the steric constraints imposed by the nodal surface 'tent'. The Fermi energy
we rediscover immediately as the corresponding energy scale $E_F=\hbar/\tau_c$. What does this imply for a critical 
fermionic state having no knowledge whatsoever about $E_F$ as required by the underlying scale invariance and as observed in various experiments? Turning the above scaling argument around, the absence of a Fermi degeneracy scale immediately implies
that the nodal surface cannot possibly carry a characteristic scale. The nodal surface constraints have to act in the same way on
all time and length scales and therefore, the nodal surface of any fermionic critical state has to be a scale invariant fractal. 
Hence, by using scaling arguments resting on the constrained path-integral, we have discovered a phenomenological principle: \emph{The collapse of the Fermi liquid at a quantum critical point as observed experimentally for instance in the heavy fermion metals is necessarily associated with a qualitative change of the nodal surface from a smooth to a fractal geometry.} This should be regarded as the most important finding reported in this Paper since it identifies the probabilistic constrained path integral 
as the mathematical framework to reconcile the workings of Fermi-Dirac statistics and scale invariance.

\section{Feynman backflow wave functions: prelimenaries}
\label{sec.feynman}

In the previous section we have convinced ourselves that a critical fermionic state is necessarily characterized by an
underlying fractal nodal surface. However, an inherent difficulty of the present approach is that in order to study nodal structures associated with non-conventional states a wave-function ansatz is required. Let us focus on a concept introduced by 
Feynman and Cohen incorporating hydrodynamical backflow effects in a quantum mechanical wave-function.\cite{feynman}

They argued that the roton in $^4$He is like a single mobile atom which is however dressed up by collective motions 
in the liquid. Helium is a nearly incompressible fluid in the hydrodynamical sense and the density in the neighborhood of 
the moving particle should be barely altered. As a consequence there has to be a backflow of other particles 
conserving the total current and leading to an enhancement of the effective mass of this quasiparticle
which can be described quantum mechanically by taking a plane-wave wavefunction $\exp(i\bk\btr_i)$ with a collective 
quasiparticle coordinate

\begin{equation}
\btr_i =  \br_i +\sum_{j(\neq i)}\eta(r_{ij})(\br_i-\br_j),
\label{bfc}
\end{equation}
where $\br_i$ are the coordinates of the bare particles, $r_{ij}=|\br_i-\br_j|$, and $\eta(r)$ a smoothly varying function 
falling off like $\sim r^{-3}$ on large distances corresponding to hydrodynamical, dipolar backflow of an incompressible fluid 
in two dimensions. Much later, it was found out that by using fermionic backflow wave functions of the form 
$\Psi\simeq\det(e^{i\bk_i\btr_j})$ for node fixing one obtains excellent variational energies for the fermionic $^3$He quantum 
fluid\cite{schmidt+81} and the homogeneous electron gas.\cite{kwon+93,holzmann+03} 

It easy to check that such slater determinants of plane-wave functions of collective backflow coordinates are indeed 
obeying Fermi-Dirac statistics since the permutation of two particles $\br_i$, $\br_j$ leads to an interchange of the 
collective coordinates $\btr_i$ and $\btr_j$ without changing the other collective coordinates and therefore to an overall 
sign change of the determinant. Therefore, as for any fermionic state the lower dimensional Pauli surface is still a
submanifold of the nodal hypersurface of fermionic backflow wavefunctions. However, due to the collectiveness build into
these wave functions one might expect the nodal hypersurfaces to be radically different from the smooth free fermion case.
This is supported by earlier work\cite{calder+95} reporting a precursor of a roughening of the nodal structure in the regime
of weak backflow.

We would like to emphasize that the hydrodynamical Feynman backflow build into the fermionic wave functions has 
nothing to do with the conventional notion of backflow in a Fermi liquid since it involves interactions between a macroscopic 
number of particles at the same time. To rationalize this we will derive an expansion of the Hamiltonian of which the backflow 
wave functions are eigenstates. In terms of collective backflow coordinates (\ref{bfc}) we are simply dealing with a gas of
free quasiparticles and the exact Hamiltonian of the system is therefore given by

\begin{equation}
\H=\sum_\bk\epsilon_\bk \ct^\dagger_\bk \ct_\bk,
\end{equation}
where we have introduced operators $\ct^\dagger_\bk$, $\ct_\bk$ creating and annihilating a backflow particle with 
momentum $\bk$, respectively. To find the representation of this Hamiltonian in terms of bare-particle operators 
$c^\dagger_\bk$, $c_\bk$ let us first derive an expansion of the $N$-particle backflow wave function in terms of free-particle
states. Using the above Fermi operators for backflow and bare particles a general relation can be written as

\begin{eqnarray}
 |\bk_1,\ldots,\bk_N\rangle_{\textrm{bf}} & = & \ct^\dagger_{\bk_1}\ldots\ct^\dagger_{\bk_N}|0\rangle\nonumber\\
& = & \int_{\bq_1,\ldots\bq_N}\Gamma_{\bq_1,\ldots,\bq_N}c^\dagger_{\bk_1+\bq_1}\ldots c^\dagger_{\bk_N+\bq_N}|0\rangle\nonumber\\
& = & \int_{\bq_1,\ldots\bq_N}\Gamma_{\bq_1,\ldots,\bq_N}\nonumber\\
& & \times|\bk_1+\bq_1,\ldots,\bk_N+\bq_N\rangle,
\label{exp}
\end{eqnarray}
where the function $\Gamma$ is defined as the Fourier transform

\begin{eqnarray}
\Gamma_{\bq_1,\ldots,\bq_N} & = & \frac{1}{V^N}\int\ud^d\br_1\cdots\ud^d\br_N \nonumber\\
& & \times\prod_i e^{-i\bq_i\br_i} e^{\sum_{j(\neq i)}f_i(\br_{ij})},
\label{ft}
\end{eqnarray}
and for abbreviation we have defined $f_i(\br_{ij})=i\eta(r_{ij})\bk_i(\br_i-\br_j)$. The general expressions (\ref{exp},\ref{ft}) 
can be further evaluated by expanding the exponentials,
\begin{eqnarray}
\prod_i e^{\sum_{j(\neq i)}f_i(\br_{ij})} & = & 1+\sum_{i,j}^{i\neq j}f_i(\br_{ij})\nonumber\\
& & +\frac 12\sum_{i,j,m,n}^{i\neq j,m\neq n}f_i(\br_{ij})f_m(\br_{mn})+\ldots,\quad
\end{eqnarray}
which leads to an expansion of the operator $\hat{A}$ which transforms the free fermion state to the backflow state,

\begin{equation}
 |\bk_1,\ldots,\bk_N\rangle_{\textrm{bf}} =  \hat{A}|\bk_1,\ldots,\bk_N\rangle.
\end{equation}
Up to second order in the backflow function $\eta(r)$ we obtain

\begin{eqnarray}
\hat{A}_0 & = & 1\nonumber\\
\hat{A}_1 & = & \sum_{i,j}^{i\neq j}\int_{\bq}\tilde{f}_i(\bq)c^\dagger_{\bk_i+\bq/2}c_{\bk_i}c^\dagger_{\bk_j-\bq/2}c_{\bk_j}\nonumber\\
\hat{A}_2 & = & \frac 12 \hat{A}_1^2\nonumber\\
& = & \frac 12\sum_{i,j,m,n}^{i\neq j,m\neq n}\int_{\bq}\int_{\bq'}  \tilde{f}_i(\bq)\tilde{f}_m(\bq')c^\dagger_{\bk_i+\bq/2}c_{\bk_i}\nonumber\\
 &&\times c^\dagger_{\bk_j-\bq/2}c_{\bk_j}c^\dagger_{\bk_m+\bq'/2}c_{\bk_m}c^\dagger_{\bk_n-\bq'/2}c_{\bk_n},
\end{eqnarray}
where we have introduced the Fourier transform

\begin{equation}
\tilde{f}_i(\bq)=\frac 1V \int\ud^d\br e^{-i\bq\br} f_i(\br).
\end{equation}
The Hamiltonian transforms under the inverse of the operator $\hat{A}$, namely

\begin{equation}
\H=\left(\hat{A}^\dagger\right)^{-1}\H_0\hat{A}^{-1},
\end{equation} 
where $\H_0=\sum_\bk\epsilon_\bk c^\dagger_\bk c_\bk$ is the Hamiltonian for the non-interacting Fermi gas. To zeroth order
(no backflow) we trivially obtain $\hat{A}=1$ and therefore $\H=\H_0$. The next corrections to the Hamiltonian in first and 
second order of the backflow function $\eta(r)$ are given by 

\begin{eqnarray}
\H_1 & = & -\hat{A}_1^\dagger\H_0-\H_0 \hat{A}_1\nonumber\\
\H_2 & = & \hat{A}_1^\dagger\H_0\hat{A}_1+\frac 12 \left(\hat{A}_1^\dagger\right)^2\H_0+\frac 12\H_0\hat{A}_1^2.
\end{eqnarray}

From the above expansion it becomes immediately clear that Feynman backflow generates a hierarchy of processes involving an increasing number of particles. To zeroth order in $\eta$, the backflow state is of course trivially identical to the free fermion one,
$\hat{A}_0=1$. Whereas the first order terms $\hat{A}_1$ are solely corresponding to pair scattering processes to second order
$\hat{A}_2$ the Hamiltonian already contains 3- and 4-body interactions. The expansion in powers of $\eta(r)$ is under 
control only if the the strength of the backflow is sufficiently weak. In this case backflow can be to a good approximation 
rationalized in terms of two-body interactions. However, when the backflow becomes strong the expansion breaks down and $n$-body interactions of arbitrary order are no longer negligible. Surely, such a state characterized by hydrodynamical backflow in a true sense involving a macroscopic number of particles cannot be obtained perturbatively since it requires an infinite number of vertex corrections. 

Anticipating that we will find a critical non-Fermi liquid state when the backflow turns hydrodynamical, it might well be that this 
gives a way a general wisdom. The Fermi liquid is known to be remarkably stable as long as one is dealing with a scaling limit
where two (quasi)particle interactions are less irrelevant than the three particle interactions and so forth. The backflow
ansatz seems to suggest that it is a necessary condition for the destruction of the Fermi liquid that the scaling flow is such that 
$n$-point interactions, with $n$ arbitrarily large, are equally marginal. 

\section{The nodal structure of backflow wave functions}
\label{sec.backflow}

\begin{figure}
\begin{centering}
\epsfig{figure=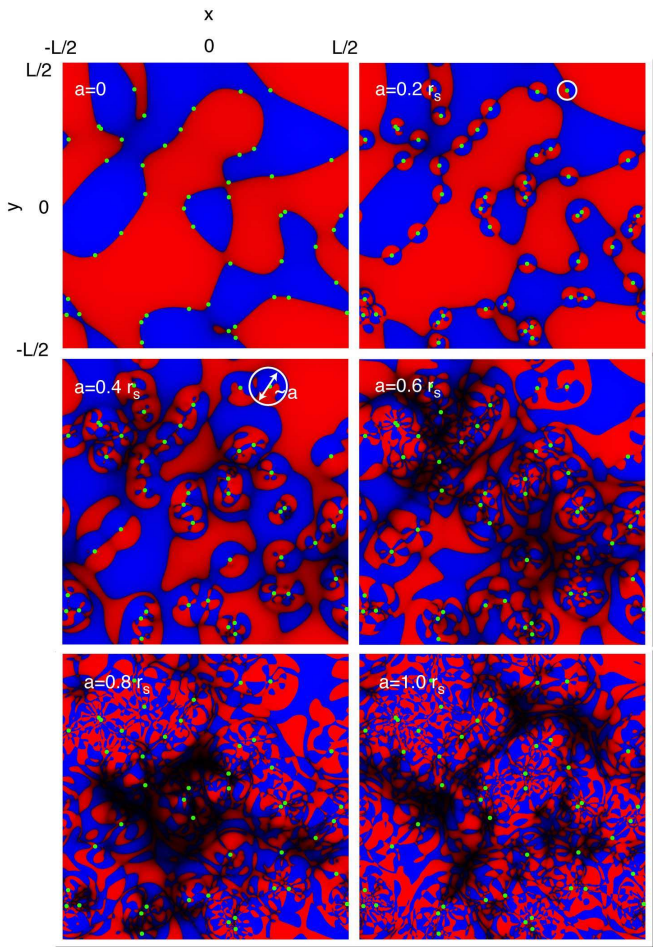,width=\linewidth}
\end{centering}
\caption{Two dimensional cuts through the nodal hypersurfaces of fermionic backflow wave-functions for $N=49$ 
particles and different values of the the backflow strength $\alpha=a/r_s$ and a small-distance cutoff $r_0/r_s=0.1$. 
The cuts are obtained in the same way as described in the caption of Fig. (\ref{freeandtent}). For $\alpha=0$ 
we recover the smooth nodal structure of free fermions. With increasing backflow strength, additional clouds 
of nodal pockets start to develop. The linear dimension of these clouds scales linearly with $\alpha$. When the 
effective backflow range $a$ becomes of the order of the inter-particle spacing $r_s$ the nodal surface 
qualitatively changes its geometry and seems to turn into a fractal.}
\label{backflow}
\end{figure}

Let us now proceed to calculate the nodal surfaces of fermionic Feynman backflow wave functions. In particular, we take 

\begin{equation}
\Psi(\br_1,\ldots,\br_N) = \mathcal{N}\det\left(e^{i\bk_i\btr_j}\right)_{i,j=1,\ldots,N}
\end{equation} 
with a set $\bk_1\ldots,\bk_N$ of momenta corresponding to a non-degenerate ground state in the free case 
(see Fig. (\ref{freeandtent}a)) and use collective backflow coordinates (\ref{bfc}) with a backflow function

\begin{equation}
\eta(r)=\frac{a^3}{r^3+r_0^3},
\label{eta}
\end{equation}
having the characteristic $r^{-3}$ tail of hydrodynamical dipolar backflow in a two-dimensional incompressible fluid. Further,
we have introduced a small distance cut-off $r_0$ and a length $a$ to make the backflow function $\eta(r)$ 
dimensionless and to control the strength of the backflow. The expansion in terms of the backflow strength should be
controlled by the dimensionless parameter $\alpha=a/r_s$ with $r_s$ the average inter-particle spacing. However, 
on the wave function level  there is no restriction to small values of $\alpha$ and we can in principle follow the evaluation of
the nodal surface into a regime where the expansion breaks down and backflow is governed by infinite number of particle correlations.

Like in the free fermion case we calculate the nodal structure on  two-dimensional cuts obtained by fixing $N-1$ particles
at random positions and searching for nodes when moving the remaining particle over the box. Due to the collectiveness
of the backflow wave functions the complexity of the problem increases significantly compared to the free case since the 
change of the coordinate of one particular particle leads to a change of all collective coordinates $\{\btr_i\}$ and we have
to recalculate the full $N\times N$ determinant for every point on the cut. For details on the node searching algorithm we
refer the reader to appendix \ref{a.nodes}.

In Fig. (\ref{backflow}) the evaluation of the nodal surface on a particular cut for $N=49$ particles with increasing backflow
strength $\alpha=a/r_s$ is shown. As a reference, in the upper left frame the nodal surface cut for free fermions corresponding 
to $\alpha=0$ is plotted. As discussed earlier the nodal surface smoothly connects the lower dimensional Pauli surface forming
pockets with a characteristic dimension of the order of the inter-particle spacing $r_s$. For small values of $\alpha$ additional 
nodal pockets start to develop for small particle separations (in the vicinity of the Pauli surface) consistent with the local 
roughening of the nodal surface reported previously,\cite{calder+95} whereas on larger scales the nodal structure looks similar 
to the free case. The size of these clouds of additional nodal pockets is much smaller than the inter-particle spacing suggesting
that backflow is basically governed by two-particle correlations as also expected from the expansion in terms of powers of 
$\eta(r)$ (see Eq. \ref{exp}). This is confirmed by the analysis of the nodal structure of two
particles subject to a mutual backflow. For small particle separations we basically observe the same local change of the nodal surface as in the $N$-particle case. These local two-particle effects have surely nothing to do with the concept of collective
hydrodynamical backflow as originally introduced by Feynman.\cite{feynman} Moreover, a decrease of the small-distance 
cut-off $r_0$ leads to the development of a local unphysical fractality in the clouds completely unrelated to collective behavior.
A closer inspection of the fermionic backflow wave functions used for node fixing in Monte-Carlo calculations shows that 
they belong to the regime of weak backflow accounting for local two-particle correlations ("exchange-correlation hole") 
in a crude way. This is further amplified by the fact that even short range backflow functions $\eta(r)$ have been 
commonly used.\cite{Cassuleras+00}

We find that the 
size of the clouds of additional nodal pockets scale linearly with $\alpha$ identifying the parameter $a$ as the effective
backflow range (see Fig. (\ref{backflow})). With increasing $\alpha$, the clouds start to overlap more and more forming clusters of interfering backflow patterns signaling that the nodal structure can no longer be understood in terms of two-particle correlations and that the backflow becomes more and more collective. 
The nodal surface seems to develop a scale invariance up to this cluster size $\xi$ which can be further amplified by decreasing the small distance cutoff $r_0$.  At the point where $\alpha$ becomes of order unity 
the backflow becomes collective involving a macroscopic number of particles. At this point $\xi$ becomes of the order
of the system size and the nodal surface seems to turn into a fractal.

\section{Fractal analysis}
\label{sec.fractal}

To demonstrate that we indeed succeeded to produce a scale invariant nodal surface, we evaluate the correlation integral
$C(r)$ which counts the number of pairs of points $\{\bX_i\}$ on the nodal surface with a separation smaller than $r$, 

\begin{equation}
C(r)= \lim_{n\to\infty}\frac{1}{n(n-1)}\sum_{i,j=1,\ldots n}^{i\neq j} \Theta(|\bX_i-\bX_j|-r),
\label{ci}
\end{equation}
where $\Theta(x)$ denotes the heavyside function, $\Theta(x)=1$ for $x\le0$ and $\Theta(x)=1$ for $x>0$. For a fractal object the 
correlation integral which is simply the integral of the point-to-point correlation function is expected to scale as a power law, $C(r)\sim r^{D_H}$ with an exponent $D_H$ very close to the Hausdorff dimension of the fractal.\cite{grassberger+83}

\begin{figure}
\begin{centering}
\epsfig{figure=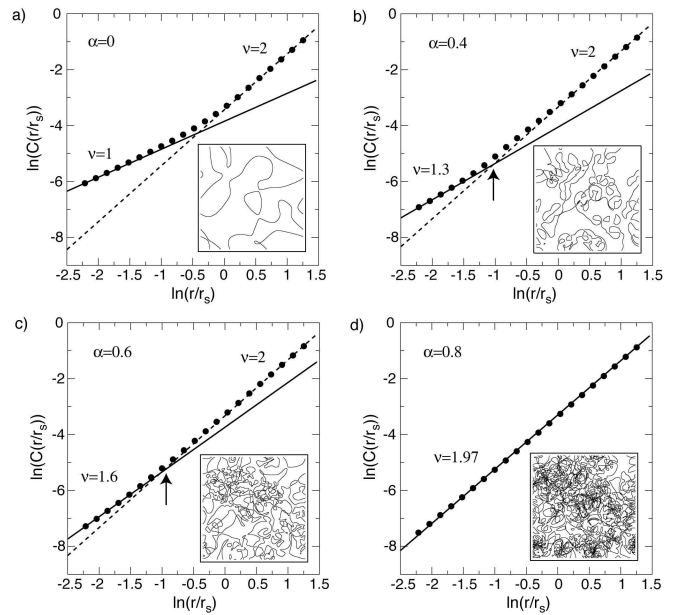,width=\linewidth}
\end{centering}
\caption{Correlation integrals $C(r)$ as a function of separation in a log-log plot for the same nodal surface cuts 
and backflow strengths $\alpha$ as used in Fig. (\ref{backflow}). The corresponding nodal surface cuts are shown as insets.
The nodal surface turns into a fractal at a critical backflow strength $\alpha_c\approx0.8$.}
\label{cdim}
\end{figure}

Of course, it is impossible to map out the full high-dimensional nodal surface. Therefore we instead perform the fractal analysis 
on various two-dimensional random cuts as shown in Fig. (\ref{backflow}). Whereas for these cut pictures we have calculated, 
solely for illustrational purposes, the values of the wave function on all points of a fine two-dimensional grid, for the fractal analysis we only need the points on the nodes which in Fig. (\ref{backflow}) correspond to the interface between positive and negative 
regions. To track down the nodes on the cut we use a triangulation method directly following the nodal lines. This 
algorithm which is described in detail in appendix \ref{a.nodes} has the virtue of omitting the calculation of the highly 
collective wave functions on too many points away from the nodes. To test the algorithm we have calculated the nodes on the 
same two-dimensional cut and for the same parameters as in Fig. (\ref{backflow}) and found a perfect agreement between 
the nodal points obtained from the triangulation method (see insets of Fig. (\ref{cdim})) and the interfaces in Fig. (\ref{backflow}).

A fractal nodal surface in full $dN$-dimensional configuration space must have a Hausdorff dimension $D_H$ between $dN-1$
which is the dimension of a regular nodal hypersurface and $dN$, the dimension of the embedding configuration space.
Therefore, in $d=2$ space dimensions one expects on a two-dimensional cut a fractal dimension $\nu=D_H/N$ in the range 
$2-1/N<\nu<2$. Hence, with increasing number of particles $N$ we expect the Hausdorff dimension $\nu$ of the nodal surface 
cut to increase and to approach $\nu=2$ in the limit $N\to\infty$. This means that for a large number of particles the nodes on the
cuts should be very close to space filling. Notice that the visual inspection of the cuts can be quite misleading regarding the 
space-filling properties of the high dimensional nodal surface.

Let us first analyze the correlation integrals (\ref{ci}) for the nodal surface cuts shown in Fig. (\ref{backflow}). 
To correct for finite-size effects we multiply $C(r)$ with a function $g(r/L)$ which normalizes the correlation integral by the 
number of available pairs of separation smaller than $r$ in a finite box $[-L/2,L/2]\times[-L/2,L/2]$. This function is given by $g(r/L)=\pi r^2/\langle A_r(x,y)\rangle_{L^2}$, where $A_r(x,y)$ denotes the part of the area of a circle with midpoint $(x,y)$ and radius $r$ lying within the box. The average $\langle.\rangle_{L^2}$ has to be taken over all points $(x,y)$ in the box.  A straightforward calculation yields for 
$r\le  L/2$

\begin{equation}
g\left(\frac rL\right)=\left[(1-\frac{8}{3\pi}\left(\frac rL\right)+\left(\frac{11}{3\pi}-1\right)\left(\frac rL\right)^2\right]^{-1}. 
\end{equation}

The evolution of the correlation integral is shown in Fig. (\ref{cdim}).
Without backflow ($\alpha=0$) where the nodes on the two dimensional cut are smooth lines with an average spacing of 
the order of the inter-particle distance $r_s$ we find $\nu=1$ for small distances as expected for a one-dimensional object and a crossover around $r_s$ to an exponent $\nu=2$ on larger scales (Fig. (\ref{cdim}a)). This crossover signals the existence of an average nodal pocket dimension $\sim r_s$: On this scale nodal lines start to see each other forming an object which looks two dimensional on larger scales. For small backflow strength $\alpha$ we introduce another scale $a=\alpha r_s$ in the system 
which we have identified with the size of additional clouds of nodal pockets. The change in the small distance behavior 
can clearly be seen in the correlation integral where we find fractality with a non-universal dimension $\nu$ up to the 
scale $a$. On smaller scales, this fractality is cut off by the parameter $r_0\ll r_s$. At the larger scale $r_s$ we again find 
a crossover to $\nu=2$ signaling the existence of an average nodal pocket dimension comparable to the free case (see Fig. (\ref{cdim}b)). Increasing $\alpha$ further, both the range of the scale invariant behavior and the fractal dimension increase 
(Fig. (\ref{cdim}c)).

At a critical value $\alpha_c\approx 0.8$ where $a$ becomes comparable to to $r_s$ both scales suddenly disappear and the 
correlation integral $C(r)$ turns into a power law up to the system size demonstrating that the nodal surface cut has become 
a scale invariant fractal with a Hausdorff dimension $\nu\approx 1.97$ very close to the space-filling dimension $d=2$
(see Fig. (\ref{cdim}d)). For values $\alpha\ge \alpha_c$ the nodal structure remains fractal without further increase of the
Hausdorff dimension $\nu$. 

We have found that with increasing backflow strength $\alpha$ the backflow clouds start to interfere forming clusters of linear dimension $\xi$ (see Fig. (\ref{backflow})). This length scale should be identified with the correlation length of scale invariant fluctuations corresponding to the crossover in $C(r)$ from scale invariant to regular behavior on larger scales as indicated 
by arrows in Figs. (\ref{cdim}b,c). Approaching the critical value $\alpha_c$, $\xi$ rapidly becomes of the order of the system size and should diverge in the limit $N\to\infty$. However, it is not possible to track down the divergence of the correlation length quantitatively since we cannot follow the crossover in a regime sufficiently close to $\alpha_c$.

\begin{figure}
\begin{centering}
\epsfig{figure=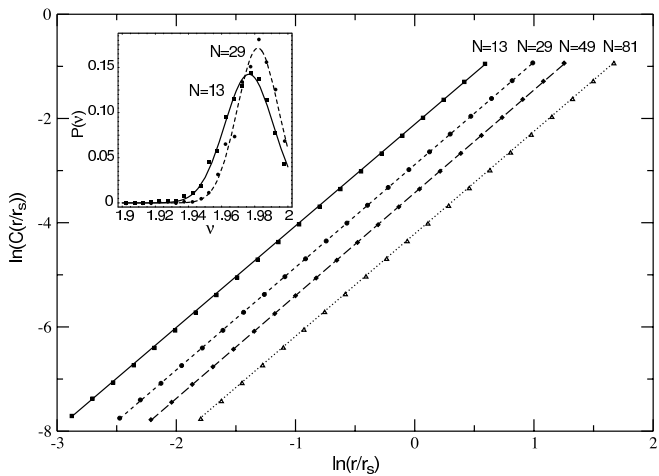,width=\linewidth}
\end{centering}
\caption{Correlation integrals $C(r)$ for random two-dimensional nodal surface cuts for different numbers of particles
and a critical backflow strength $\alpha_c=0.8$. In all cases we find scale-invariant behavior and Hausdorff dimensions slightly
below 2. In the inset we plotted the distributions $P(\nu)$ of the Hausdorff dimensions obtained from a large number of random two dimensional cuts for $N=13$ and $N=29$ particles, 
respectively.}
\label{frac}
\end{figure}

In Fig. (\ref{frac}) we have plotted the correlation integrals $C(r)$ for different numbers of particles at a fixed density
$n=N/V\sim r_s^{-2}$ taking the critical backflow strength $\alpha_c=0.8$. We find power-law behavior over approximately three decades for the biggest system and exponents $\nu$ slightly below 2 in all cases. Further, we observe a small increase 
of the fractal dimension consistent with the inequality $2-1/N<\nu<2$. For smaller systems it is possible to calculate 
$C(r)$ for a large number of random two-dimensional cuts. The resulting distributions $P(\nu)$ of the Hausdorff 
dimensions is shown in the inset of Fig. (\ref{frac}) for $N=13$ and $N=29$ particles, respectively. For the larger number of 
particles the Gaussian distribution becomes narrower and shifts towards $\nu=2$.  From the mean $\bar{\nu}$ and the width of the distribution $P_N(\nu)$ we estimate for the Hausdorff dimension of the nodal surface in full $dN$-dimensional configuration space
$D_H=N\bar{\nu}=2N-1+R_H$ with $R_H=0.6\pm0.3$ and $R_H=0.5\pm0.3$ for $N=13$ and $N=29$, respectively.

\section{Momentum distribution and effective mass divergence}
\label{sec.mass}

We have demonstrated that fermionic Feynman backflow wave functions exhibit very rich nodal structures and that
by increasing the backflow strength $\alpha$ it is possible to continuously tune the nodal surface from the smooth one 
of the free gas to a scale invariant fractal. On the other hand, we showed that the Fermi degeneracy scale has a simple 
geometrical meaning:  it has a one-to-one correspondence to an average nodal pocket dimension. This scale disappears 
as we approach the critical backflow strength. Therefore, we succeeded to produce a critical fermionic state of matter 
lacking a Fermi degeneracy scale! 

Experimentally, the collapse of a Fermi-liquid state towards a quantum critical point has been observed in a spectacular 
fashion in the heavy-fermion metals. From Hall resistivity measurements\cite{paschen} it has been concluded that the 
Fermi surface undergoes a discontinuous jump at the quantum critical point signaling that there is no sense of a Fermi surface underlying the critical state. How does the system gets rid of the Fermi energy scale?  The quasiparticles in the Fermi liquids on both sides of the transition are characterized by an effective mass $m^*\sim 1/E_F$, which diverges as a power law as one 
approaches the critical point as for example clearly seen in measurements of the linear specific heat coefficient $C(T)/T\sim m^*$.
\cite{custers} A theoretical understanding how the quasiparticles get heavier and heavier and finally completely 
die is completely lacking and it remains the question to what extend collective backflow wave functions can account for such a mysterious behavior.  

The quasiparticle effective mass $m^*$ is per definition given by

\begin{equation}
\frac{m^*}{m}=\left.\frac{m}{k}\frac{\ud}{\ud k}\epsilon(\bk)\right|_{k=k_F},
\end{equation}
with $m$ the bare electron mass and can be expressed by use of the Dyson equation\cite{Dyson49,Abrikosov+} in terms of 
the real part of the quasiparticle self energy $\Sigma(\bk,\omega)$ as

\begin{equation}
\frac{m^*}{m}=\left.\frac{1-\frac{\partial}{\partial \omega} \textrm{Re}\Sigma(\bk,\omega)}{1+\frac{m}{k}\frac{\partial}{\partial k} \textrm{Re}\Sigma(\bk,\omega)}\right|_{k=k_F,\omega=0}.
\end{equation}
Under the conventional assumption that the momentum dependence of the real part of the self energy remains non-singular
a divergence of the quasiparticle mass $m^*$ can be directly related to a vanishing quasiparticle pole strength

\begin{equation}
Z=\left.\left(1-\frac{\partial}{\partial \omega} \textrm{Re}\Sigma(\bk,\omega)\right)^{-1}\right|_{k=k_F,\omega=0},
\end{equation}
which defines the size of the effective Fermi surface discontinuity and the disappearance of the 
discontinuity $Z$ in the quasiparticle momentum distribution $n(\bk)$ would correspond to an effective mass divergence
$m^*/m\sim 1/Z$. 

Since $Z$ can be derived from the $n(\bk)$ jump, we calculate the single-particle momentum distribution of backflow wave functions to establish a connection between an effective mass divergence and the disappearance of a scale in the nodal hypersurface. From the ground-state wave function the single-particle momentum distribution at zero temperature can be calculated as a Fourier transform of the reduced one-body density matrix (\ref{nr}),

\begin{eqnarray}
n(\bk) & = &  (2\pi)^{-2}\int\ud\br\int\ud\bR e^{i\bk\br}\Psi^*(\br_1,\br_2,\dots,\br_N)\nonumber\\
& & \times\Psi(\br_1+\br,\br_2,\ldots,\br_N).
\label{nkeq}
\end{eqnarray}
For the backflow wave functions this high dimensional integral cannot be calculated analytically since the bare coordinates
enter via the transformations to collective coordinates (\ref{bfc}) in a highly non-trivial way and an expansion in terms of 
the backflow strength $\alpha$ (\ref{exp}) breaks down as we approach the critical value where the backflow becomes highly collective involving a macroscopic number of particles at the same time. Of course numerically such integrals in $D=dN+d$ 
dimensions can only be calculated by means of Monte Carlo integration. The basic idea is very simple and is based on the
central limit theorem. Let $\{\bx_1,\ldots,\bx_n\}$ be a set of uncorrelated points in a very high dimensional space which are 
distributed according to a probability distribution $P(\bx)$ fulfilling the requirements $P(\bx)\ge0$ and $\int \ud\bx P(\bx)=1$. 
Lets introduce a new random variable $Y_f=[f(\bx_1)+\ldots+f(\bx_n)]/n$ where $f$ is some arbitrary real-valued function with
mean $\mu_f$ and variance $\sigma_f^2$ given by

\begin{eqnarray}
\mu_f & = & \langle f(\bx)\rangle_{P}\nonumber\\
& = &  \int \ud \bx f(\bx)P(\bx),\\
\label{mean}
\sigma_f^2 & = & \left\langle \left( f(\bx)-\langle f(\bx)\rangle\right)^2 \right\rangle_{P}\nonumber\\
& = &  \int \ud \bx (f(\bx)-\mu_f)^2 P(\bx).
\label{var}
\end{eqnarray}
Then it can be shown that under rather general conditions\cite{Feller} that the central limit theorem applies and that for large enough $n$ the variable $Y_f$ is normally distributed with mean $\mu_f$ and standard deviation $\sigma_f/\sqrt{n}$, irrespective 
of the dimension. These ideas can easily be applied to high dimensional integrals like for the momentum distribution we want to calculate. To establish a connection with Eq. (\ref{mean}) we simply rewrite the integral  by introducing
a probability distribution, $I= \int \ud \bx g(\bx) = \int \ud \bx f(\bx)P(\bx)$ with $f(\bx)=g(\bx)/P(\bx)$. The integral $I$ can then be estimated by sampling a large number $n$ of points $x_i$ according to the probability distribution $P(\bx)$ as

\begin{equation}
I=\int \ud \bx g(\bx)=\left\langle \frac{g(\bx)}{P(\bx)}\right\rangle_P   \approx \frac 1n \sum_{i=1}^n \frac{g(\bx_i)}{P(\bx_i)},
\end{equation}
whereas the variance of the estimate of the integral is given by

\begin{equation}
\frac{\sigma_f^2}{n}\approx\frac{1}{n^2}\sum_{i=1}^n\left[\frac{g(\bx_i)}{P(\bx_i)}-\frac 1n \sum_{j=1}^{n} \frac{g(\bx_j)}{P(\bx_j)}\right]^2,
\label{error}
\end{equation}
from which we obtain $\pm \sigma_f/\sqrt{n}$ as an estimate for the size of the error bar on the computed value of $I$. Obviously,
for a given sample size the variance depends significantly on the choice of the probability function $P(\bx)$. In case of
the integral $n(\bk)$ (\ref{nkeq}) we have tried different distributions and found the best convergence for the choice

\begin{equation}
P(\bx)=P(\bR,\br)=|\Psi(\bR)|^2,
\label{pdist}
\end{equation}
giving for every $\bk$ point the value of $n(\bk)$ as the sample average
 
\begin{equation}
n(\bk)=\left\langle e^{i\bk\br}\frac{\Psi(\br_1+\br,\br_2,\ldots,\br_N)}{\Psi(\br_1,\br_2,\ldots,\br_N)}\right\rangle_{P(\bR,\br)}. 
\end{equation}

To sample points $\bR=(\br_1,\ldots,\br_N)$ in $2N$-dimensional configuration space according to the density profiles $\rho(\bR)=|\Psi(\bR)|^2$ of backflow wave functions for different particle numbers and parameters $a$ and $r_0$ (see Eq. (\ref{eta})) we use a standard Metropolis rejection algorithm,\cite{Metropolis} which has the great advantage that it allows an arbitrarily complex distribution to be sampled in a straightforward way without knowledge of its normalization. For a very detailed description of the Metropolis algorithm but also for an overview on quantum Monte Carlo methods  in general we refer the reader to Ref. \onlinecite{Foulkes+01}. Surely the density profile $\rho(\bR)$ becomes very craggy as we approach the critical backflow strength 
$\alpha_c\approx 0.8$ where the zeros of the density form a scale invariant
fractal.   Since the probability distribution  of choice (\ref{pdist}) does not depend on the coordinates $\br$ we just pick these points on the plane randomly.

\begin{figure}
\begin{centering}
\epsfig{figure=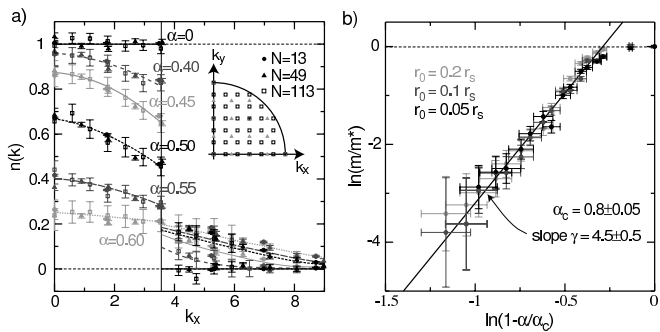,width=\linewidth}
\end{centering}
\caption{a) Momentum distribution $n(\bk)$ along $k_x$ for different backflow 
strengths $\alpha$ for $N=13$ (filled circles), $N=49$ (filled triangles), and $N=113$ (open squares) particles 
at a fixed density. Here, a relatively large cut-off $r_0=0.4 r_s$ has been used. b) Inverse effective 
quasiparticle mass $m/m^*$ as a function of $\alpha$ for $r_0/r_s=0.2, 0.1, 0.05$. $m/m^*$ has been obtained 
from an interpolation of the discontinuity of $n(k)$. Due to numerical convergence problems we were not able to 
follow the behavior for $\alpha >0.6$. Taking the value $\alpha_c\approx 0.8$ extracted from the correlation 
integral analysis the behavior of $n(k)$ up to $\alpha=0.6$ suggests a strong mass divergence $m^*/m\sim(1-
\alpha/\alpha_c)^{-\gamma}$ with $\gamma=4.5\pm0.5$. }
\label{nk}
\end{figure}

The resulting normalized momentum distributions along the $k_x$ direction for different values of $\alpha$ are shown in Fig. (\ref{nk}a) where the error bars $\pm \sigma_f/\sqrt{n}$ are obtained from Eq. (\ref{error}). 
We have used a relatively large small-distance cut-off $r_0=0.4r_s$ to suppress the fractality of the nodal surfaces on smallest scales leading to a highly oscillatory behavior of the integrands in the ultraviolet. For finite size scaling, we have evaluated $n(k)$
for each value of $\alpha$ for various numbers of particles at a fixed density leading to an
increase the momentum resolution $\Delta k=2\pi/L$. For different numbers of particles we find a consistent 
interpolation of $n(\bk)$ for all values of $\alpha$ indicating that the finite size scaling is well behaved. 
In the inset of Fig. (\ref{nk}a) the corresponding set of momenta entering the wave functions is shown. To avoid confusion, we would like to emphasize that these momentum states within the Fermi 
sphere are fully occupied only in the free fermion case. As seen in the expansion (\ref{exp}) of the backflow wave functions
in terms of free particle states backflow leads to a hierarchy of scattering processes leading to a mixing of all kind of excited 
free particle states and must therefore result in a drastic change of the quasiparticle momentum distribution. 

For $\alpha=0$ we obtain the discrete step function of the free Fermi gas. With increasing backflow strength $\alpha$ the discontinuity $Z$ decreases continuously consistent with the picture that the bare particles get dressed with backflow clouds 
(see Fig. (\ref{backflow})) leading to an enhancement of the quasiparticle effective mass. In the regime of small backflow 
we find the numerical convergence to be extremely fast even for large systems. For larger values of $\alpha$ where the
backflow clouds start to overlap (Fig. (\ref{backflow})), $Z$ starts to decrease very rapidly.
Approaching the critical value $\alpha_c\approx0.8$ where the nodal hypersurface turns into a fractal, for an increasing particle number $N$ the signal to noise ratio goes down significantly. This critical slowing down in the numerics indicates that 
we are facing the singular limit of infinite number of particle correlations. For $\alpha>0.6$ it becomes impossible to extract the 
discontinuity since it becomes smaller than the numerical resolution. Qualitatively, a disappearance of the discontinuity at the
the critical value $\alpha_c\approx 0.8$ where the nodal surface turns into a fractal seems  consistent with the numerical 
data.

To extract the form of the effective mass divergence $m^*/m\sim 1/Z$ as suggested by the disappearing of the discontinuity
$Z$ we use the value $\alpha_c=0.8$ extracted form the correlation integrals. We find that the behavior of $1/Z$ up to 
$\alpha=0.6$ is indeed consistent with a strong power-law divergence of the quasiparticle mass 

\begin{equation} 
\frac{m^*}{m}\sim\left(1-\frac{\alpha}{\alpha_c}\right)^{-\gamma}
\end{equation} 
with an exponent $\gamma=4.5\pm0.5$ (see Fig. (\ref{nk}b)). Smaller values of the cut-off $r_0$ lead to a more rapid enhancement
of the quasiparticle mass for very small values of $\alpha$ but do not influence the form of the effective mass divergence. This
indicates that the effective mass divergence going hand in hand with the emergence of scale invariance of the nodal structure on large scales is driven 
by collective backflow correlations involving infinite particle interactions, or in diagrammatic language infinite order vertex 'corrections'.

\section{Conclusion}
\label{sec.disc}

In summary, employing the constrained path-integral formalism we have delivered here proof of principle that 
fermion statistics and the emergent scale invariance underlying the critical state can be reconciled. 
We perceive it as highly profound that the workings of fermion statistics in interacting many particle systems 
can be encoded in a geometrical structure (the nodal surface) which in turn can be married with the symmetry 
of scale invariance to yield a description of fermionic quantum critical states. Phenomenologically, the physics of 
fermion systems can be viewed as 'boson dynamics times nodal surface geometry' and on this level it is in principle 
a tractable problem to impose a fractal nodal surface, to subsequently compute propagators and thermodynamical 
properties. Of course, one anticipates that the fractal dimension of the nodal surface enters thermodynamic exponents
and sets an anomalous dimension in the propagators. For the latter we have already found signatures from the 
diffusive behavior imposed by the fractal nodal surface generated by backflow. Instead of conventional Gaussian 
diffusion we find super-diffusive behavior corresponding to single-particle propagators acquiring a Levy-flight form.
In principle, there is hope for a generalization of Kadanoff-type scaling relations or of Wilsonian renormalization group 
treatment of fermionic systems around such non-Gaussian quantum critical  points.  

Also the competition with superconductivity can be studied: 
it has been demonstrated that the nodal structure associated with BCS superconductors is subtly different from 
that of Fermi gasses\cite{akkineni+06}, and it would be quite interesting to find out how this affair would work out 
starting from a fermionic critical state. Naively, one expects that a fractal nodal surface imposing constraints on the
dynamics on all length and time scales would lead to a drastic enhancement of pairing which might be the reason why
quite generically instabilities towards superconducting order in the vicinity of fermionic quantum critical points have been
observed.

We have used here the backflow wave function just as a device to generate a fractal nodal surface, leaving open 
what the actual microscopic conditions are, causing criticality in the physical systems. 
Backflow is associated with the effective incompressible nature of the fluid flow and this has an interesting resemblance 
with the microscopy of the strongly correlated electron systems:  in one or the other way, the Mott insulator is close
by when fermionic quantum criticality is observed and 'Mottness' renders electron systems
to become incompressible. In fact, the physics associated with backflow is quite similar
to the interpretation given by Anderson of the 'strange metal' as a Gutzwiller projected fermion 
systems,\cite{anderson+06} in the sense that both approaches are based on a hidden quasiparticle 
picture and involve singular transformations of the Fermi liquid state. It needs further investigation 
in what sense the nodal structures associated with Gutzwiller-projected wave functions are scale
invariant and resemble critical states of fermionic matter. 

\section{Acknowledgments}

We acknowledge insightful discussions with S.~C. Zhang, A.~V. Balatsky, L. Mitas, and D.~M. Ceperley.
This research was supported by the "Nederlandse organisatie voor Wetenschappelijk Onderzoek"  (NWO) and 
by the "Stichting voor Fundamenteel Onderzoek der Materie" (FOM).                                                                        

\appendix

\section{Node searching}
\label{a.nodes}

To find the nodes of the free fermion and Feynman backflow wave functions we have to track the sign changes of $\Psi(\bR)$
on a numerical grid where the nodal structure has to look smooth on the scale of the grid size $\epsilon$. Obviously, it is not 
possible to map out the full high-dimensional nodal hypersurface in $dN$-dimensional configuration space. Instead, we calculate
the nodal structures on random two dimensional cuts, obtained by keeping $N-1$ particles  fixed at random positions $\br_2,\ldots,\br_N$, and tracking down the zeros of the wave function when moving the remaining particle $\br_1$ over the system $[-L/2,L/2]
\times [-L/2,L/2]$,

\begin{equation}
\Omega_{\br_2,\ldots,\br_N}=\left\{\br_1|\Psi(\br_1,\br_2,\ldots,\br_N)=0  \right\}.
\end{equation}
If the nodal surface is not fractal this defines a 1-dimensional hypersurface in the $(d=2)$-dimensional configuration space of the
particle $\br_1$. To calculate the values of backflow wave functions on a certain grid point we have to evaluate an ($N\times N$)-dimensional determinant, $\det \bA$ with $a_{ij}=\exp(i \bk_i\tilde{\br}_j)$ with the collective backflow coordinates given in Eq. (\ref{bfc}). This is done by using the LU-decomposition $\bA=\bL\bU$ where $\bL$ and $\bU$ are lower and upper triangular matrices of
dimension $N\times N$, respectively. This decomposition is unique if we require $l_{ii}=1$ for the diagonal elements of $\bL$. In this
case, the determinant is given by

\begin{equation}
\det \bA=\det(\bL\cdot\bU)=\det(\bL)\det(\bU)=\prod_{i=1,\ldots,N}u_{ii}.
\end{equation}  
It is easy to check that one can adjust the overall pre-factor of the wave function to make it real valued. This is a necessary condition
to obtain the nodes by the sign changes of $\Psi$.

It becomes immediately clear why the collective backflow makes it is much harder to calculate the nodes. In the case of free fermions
($\alpha=0$) we have $\tilde{\br}_i=\br_i$ and by moving the particle $\br_1$ over the cut we change only the first row of the matrix
$\bA$. Therefore we can use the expansion

\begin{equation}
 \det(\bA)=\sum_{i=1,\ldots,N}(-1)^{1+i}a_{1i}\det(\bA_{1i}),
 \end{equation}
 where $\bA_{1i}$ denotes the $(N-1)\times(N-1)$ matrix obtained by dropping the first row and the $i$-th column of the matrix $\bA$.
 Only the prefactors $a_{1i}$ depend on the coordinate $\br_1$ whereas the sub-determinants have to be calculated only once for 
 a particular cut. Including backflow, the change of the position $\br_1$ of the first particle leads to a change of all collective
 coordinates $\{\tilde{\br}_i\}$ at the same time and the above expansion turns out to be useless. Instead, on every grid point we 
 have to recalculate all collective coordinates and the full $N\times N$ determinant.

\begin{figure}
\begin{centering}
\vspace{0.5cm}
\epsfig{figure=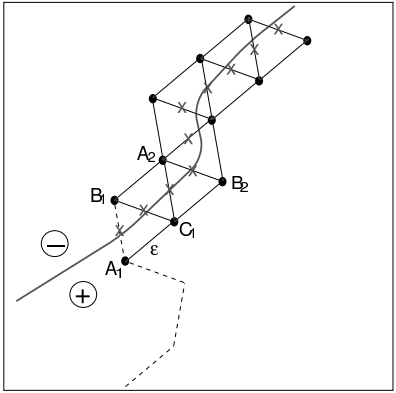,width=0.6\linewidth}
\end{centering}
\caption{Illustration of the algorithm following a nodal line by a triangulation method.}
\label{tri}
\end{figure}

To track the sign changes on the cut we use two different algorithms. The first one is based on a triangulation in the vicinity of
the nodal lines (see Fig. (\ref{tri})) avoiding the calculation of the wave function on too many points away from the nodes. A first
point on the node is found by following a random walk till we encounter a sign change, lets say between the points $A_1$ and 
$B_1$. We now initialize our triangulation procedure by choosing one of the two possibilities of completing the side $A_1B_1$ to
an equilateral triangle $\Delta_1=A_1B_1C_1$. If the nodal structure is smooth on the scale $\epsilon$ of the numerical grid
the nodal line has leave the triangle through one of the two new sides $B_1C_1$ or $A_1C_1$, in our example it crosses
the side $B_1C_1$ since $\Psi(B_1)\Psi(C_1)<0$ whereas $\Psi(A_1)\Psi(C_1)>0$. We now take the mirror image $A_2$ of the point $A_1$ with respect to the side $B_1C_1$ to obtain the next triangle $\Delta_2=B_1C_1A_2$ and repeat the procedure. Using this triangulation we directly follow the nodal line.

Obviously, the above procedure breaks down when a triangle is penetrated by two nodal lines or when a line leaves the triangle 
through the same side it had entered. Therefore, the grid size $\epsilon$ has to be sufficiently small. In particular, even for small backflow the nodal surfaces develop a local fractality (see Fig. (\ref{cdim})) which is suppressed on smallest scales
by the UV cut-off $r_0$. Hence we have to fulfill the requirement $\epsilon\ll r_0$ to not run into problems.

For illustrational purposes it is desirable to calculate the wave function on all points of a two dimensional $n\times n$ grid where
$\epsilon=L/n$. This method has been used for Fig. (\ref{backflow}) where the nodal lines correspond to the interface between
positive (red) and negative (blue) regions. The absolute values of $\Psi$ are encoded in the color shading.

\end{document}